\documentclass[
]{ceurart}

\usepackage{listings}
\usepackage{subcaption}
\begin{document}

\copyrightyear{2024}
\copyrightclause{Copyright for this paper by its authors.
  Use permitted under Creative Commons License Attribution 4.0
  International (CC BY 4.0). CEUR Workshop Proceedings (CEUR-WS.org)}

\conference{The 2nd Music Recommender Workshop (@RecSys),
  October 14, 2024, Bari, Italy}

\title{Towards Leveraging Contrastively Pretrained Neural Audio Embeddings for Recommender Tasks}

%\tnotetext[1]{You can use this document as the template for preparing your
%  publication. We recommend using the latest version of the ceurart style.}

\author[1]{Florian Grötschla}[%
orcid=0009-0004-1509-174X,
email=fgroetschla@ethz.ch
]
%\cormark[1]
\address[1]{ETH Zurich, Switzerland}

\author[1]{Luca Strässle}[%
orcid=0009-0002-5264-162X,
email=lucastr@ethz.ch
% email=luca.straessle@gess.ethz.ch
]

\author[1]{Luca A. Lanzendörfer}[%
orcid=0009-0009-5953-7842,
email=lanzendoerfer@ethz.ch
]

\author[1]{Roger Wattenhofer}[%
orcid=0000-0002-6339-3134,
email=wattenhofer@ethz.ch
]

\begin{abstract}
  Music recommender systems frequently utilize network-based models to capture relationships between music pieces, artists, and users. Although these relationships provide valuable insights for predictions, new music pieces or artists often face the cold-start problem due to insufficient initial information. To address this, one can extract content-based information directly from the music to enhance collaborative-filtering-based methods. While previous approaches have relied on hand-crafted audio features for this purpose, we explore the use of contrastively pretrained neural audio embedding models, which offer a richer and more nuanced representation of music. Our experiments demonstrate that neural embeddings, particularly those generated with the Contrastive Language-Audio Pretraining (CLAP) model, present a promising approach to enhancing music recommendation tasks within graph-based frameworks.
\end{abstract}


\maketitle

\section{Introduction}

Music and artist recommendations have become a cornerstone of streaming services, profoundly influencing how users discover and engage with music. Algorithmically generated playlists, tailored to individual tastes, are integral to the listening experience, enabling users to find music that suits their mood and environment, as well as discover new artists. For artists, inclusion in these playlists can significantly boost their listener base, while exclusion poses challenges for discovery. Music recommendation systems can be broadly categorized into collaborative filtering-based approaches~\cite{sarwar2001item} and content-based approaches~\cite{pazzani2007content}. Collaborative filtering leverages relational data, capturing relationships between artists or tracks from manually curated similarities, tags, and user listening behavior. Content-based approaches utilize descriptive data to encapsulate the essence of an artist's music, representing attributes like melody, harmony, and rhythm. Hybrid recommender systems~\cite{burke2002hybrid, adomavicius2005toward} combine both types of data to enhance recommendation quality.
In recent years, contrastive learning approaches have gained traction for their effectiveness in representing various types of data~\cite{chen2020simple, radford2021learning}. One such model, Contrastive Language-Audio Pretraining (CLAP)~\cite{wu2023large}, maps text and audio into a joint multi-modal space, offering a novel method for representing music. Our work explores the utility of CLAP representations as descriptive data in music recommendation systems.

As a proof-of-concept, we examine a graph-based artist-relationship prediction task, where additional musical information has previously enhanced model performance~\cite{korzeniowski2021artist}. The goal is to predict relationships between previously unseen artists using the attached information. By varying this information and incorporating CLAP embeddings, we evaluate its utility in a controlled environment and benchmark the effectiveness of different representations.

\section{Related Work}

\paragraph{Artist Similarity with Graph Neural Networks.}
Graph Neural Networks (GNNs)~\cite{scarselli2008graph} extend deep learning techniques to graph-structured data, addressing the limitations of traditional neural networks that require structured inputs. GNNs operate on graphs defined by nodes and edges, leveraging message passing to aggregate and update node information based on their neighbors. This approach has shown success in tasks such as node classification, edge prediction, and graph classification~\cite{wu2020comprehensive}. GNNs lend themselves to music recommender tasks as they can encode the structural, relational information together with additional features~\cite{oramas2016sound, weng2022graph}.

The study by \citet{korzeniowski2021artist} introduces the OLGA dataset, which includes artist relations from AllMusic\footnote{\url{https://www.allmusic.com/}} and audio features from AcousticBrainz~\cite{bogdanov2019acousticbrainz}. Their GNN architecture combines graph convolution layers with fully connected layers and was trained with a triplet loss. Performance evaluations on an artist similarity task demonstrated that incorporating graph layers and meaningful artist features significantly improved prediction accuracy over using deep neural networks alone.

\paragraph{Neural Embeddings for Recommender Tasks.}
Various methods have been explored for music similarity detection. Previous approaches used a graph autoencoder to learn latent representations in an artist graph~\cite{salha2021cold}, or leveraging a Siamese DCNN model for feature extraction and genre classification~\cite{park2017representation}. \citet{oramas2017deep} use CNNs to extract music information, which, in contrast to our work, can not benefit from contrastive pertaining. Furthermore, hybrid recommendation systems using GNNs have been applied in other domains, such as predicting anime recommendations by combining user-anime interaction graphs with BERT embeddings~\cite{javaji2023hybrid}.

Contrastive Language-Audio Pretraining (CLAP)~\cite{wu2023large} learns the (dis)similarity between audio and text through contrastive learning, mapping both modalities into a joint multimodal space. Through the contrastive learning approach, even the audio embeddings alone maintain semantic information, making it suitable for tasks such as music recommendation and artist similarity.

\section{Neural Audio Embeddings for Artist Relationships}

We investigate an established artist similarity task similar to the OLGA dataset to evaluate the effectiveness of neural audio embeddings over classical audio features in music recommendation tasks. This dataset comprises a large graph of artists, and the performance of our model is assessed based on its ability to predict new relationships between previously unseen artists, represented as nodes within the graph. Each node is annotated with features extracted from the music produced by the respective artist.
Previous research demonstrated that incorporating musical information significantly improves model performance~\cite{korzeniowski2021artist}. We extend this analysis by extracting CLAP embeddings from the music and comparing their effectiveness against other feature sets. Our goal is to determine if CLAP embeddings provide better representations.

\subsection{Experimental Setup}

Our setup is inspired by the approach of \citet{korzeniowski2021artist} on OLGA, where artists are represented as connected nodes based on their relationships described in AllMusic. Following the same methodology, we create an updated version of the original dataset. This allows us to ensure that the song for which we extract features from AcousticBrainz is consistent with the song for which we create CLAP embeddings. 
% This is otherwise hard to achieve based on the information from the original dataset and the fact that the provided features are already averaged over several songs. 
We start with the same set of artists and collect additional information during preprocessing, specifically the categorical features for moods and themes of an artist, which we use during evaluation. Low-level music features for songs were retrieved from AcousticBrainz, and CLAP embeddings were computed using the LAION CLAP model from tracks on YouTube.
In contrast to the original OLGA dataset, we only use one song per artist and do not aggregate the features over multiple songs. Due to constantly changing information on AllMusic, some artists without connections to other artists or missing matches on MusicBrainz or AcousticBrainz had to be dropped. Overall, this reduced the total number of artists from 17,673 in the original to 16,864 in our version. We reuse the split allocation of the OLGA dataset, which is possible since every artist in our dataset is present in the OLGA dataset as well. This resulted in 13,489 artists in the training, 1,679 artists in the validation, and 1,696 artists in the test split.
We utilize the same loss functions and GNN backbone as proposed by \citet{korzeniowski2021artist}, but with a uniform sampling based on triplets instead of distance-weighted sampling. More specifically, we employed the triplet loss, finding that using both endpoints as anchors performed better than randomly selecting one endpoint. Euclidean distance was used for the loss, and the Normalized Discounted Cumulative Gain (NDCG) serves for the evaluation. 
For the graph neural network layers, we experimented with \textsc{SAGE}~\cite{hamilton2017inductive}, \textsc{GatedGCN}~\cite{dwivedi2023benchmarking}, and \textsc{GIN}~\cite{xu2018powerful}, with \textsc{SAGE} demonstrating the best performance. 

We vary two primary aspects in our experiments: the number of graph layers and the node features. The number of graph layers ranges from zero to four and is varied to assess the contribution that the graph topology can make to the task. With zero graph layers, the architecture only utilizes an MLP to make predictions and does not consider the graph topology, thus serving as a baseline for models that use GNN layers. As the number of graph layers increases, nodes can aggregate information from a larger neighborhood, enhancing the model's capacity to learn from the graph structure. For node features, we use random features as a baseline and experimented with AcousticBrainz features, CLAP features, and Moods-Themes features. We also test combinations of these non-random features.

\subsection{Results}
\begin{figure}
    \centering
    \begin{subfigure}{.48\textwidth}
        \centering
        \includegraphics[width=\linewidth]{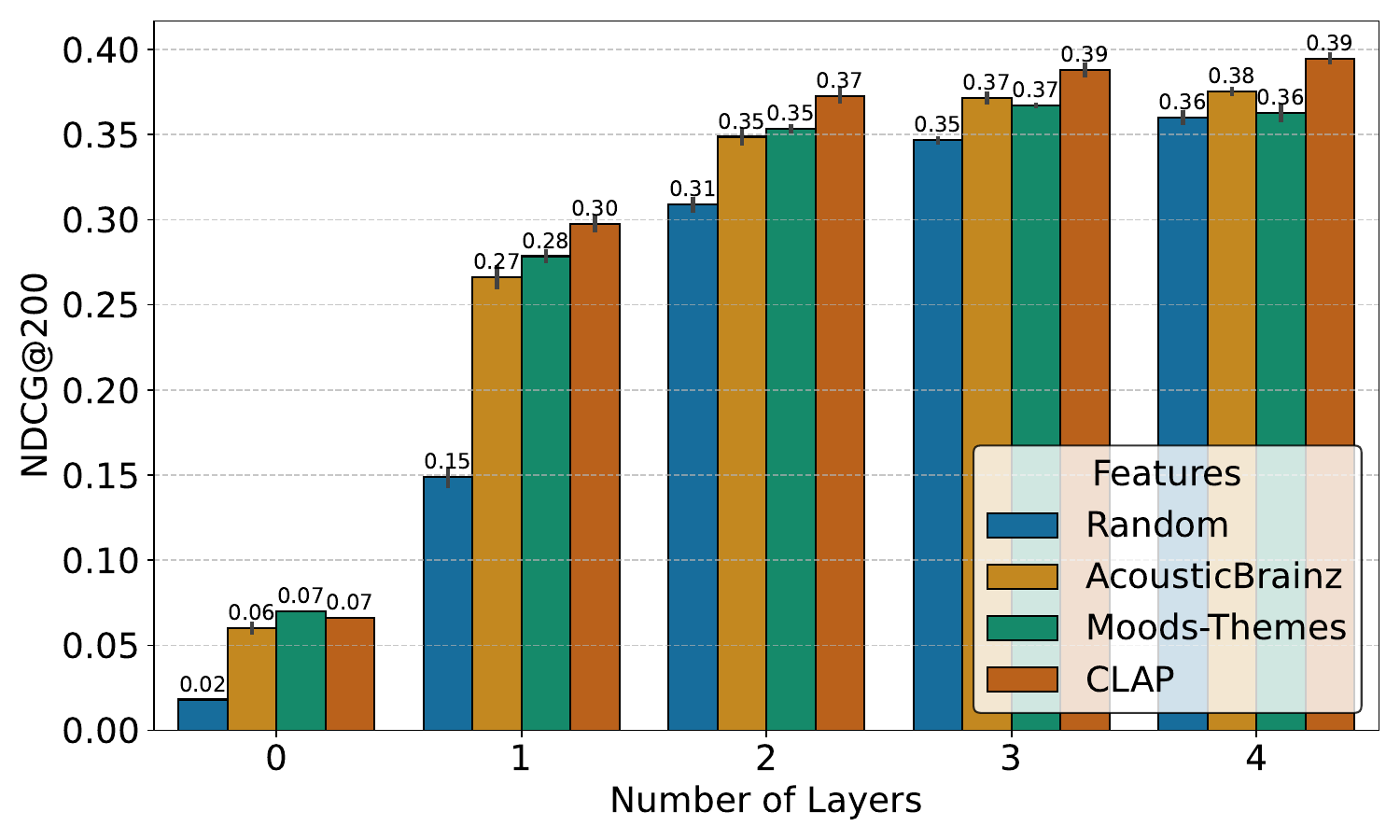}
        \caption{Comparison of CLAP features with Random, Moods-Themes, and AcousticBrainz features. CLAP outperforms all other features when used with enough layers.}
        \label{fig:results_a}
    \label{fig:enter-label}
    \end{subfigure}
    \hspace{0.1cm}
    \begin{subfigure}{.48\textwidth}
        \centering
        \includegraphics[width=\linewidth]{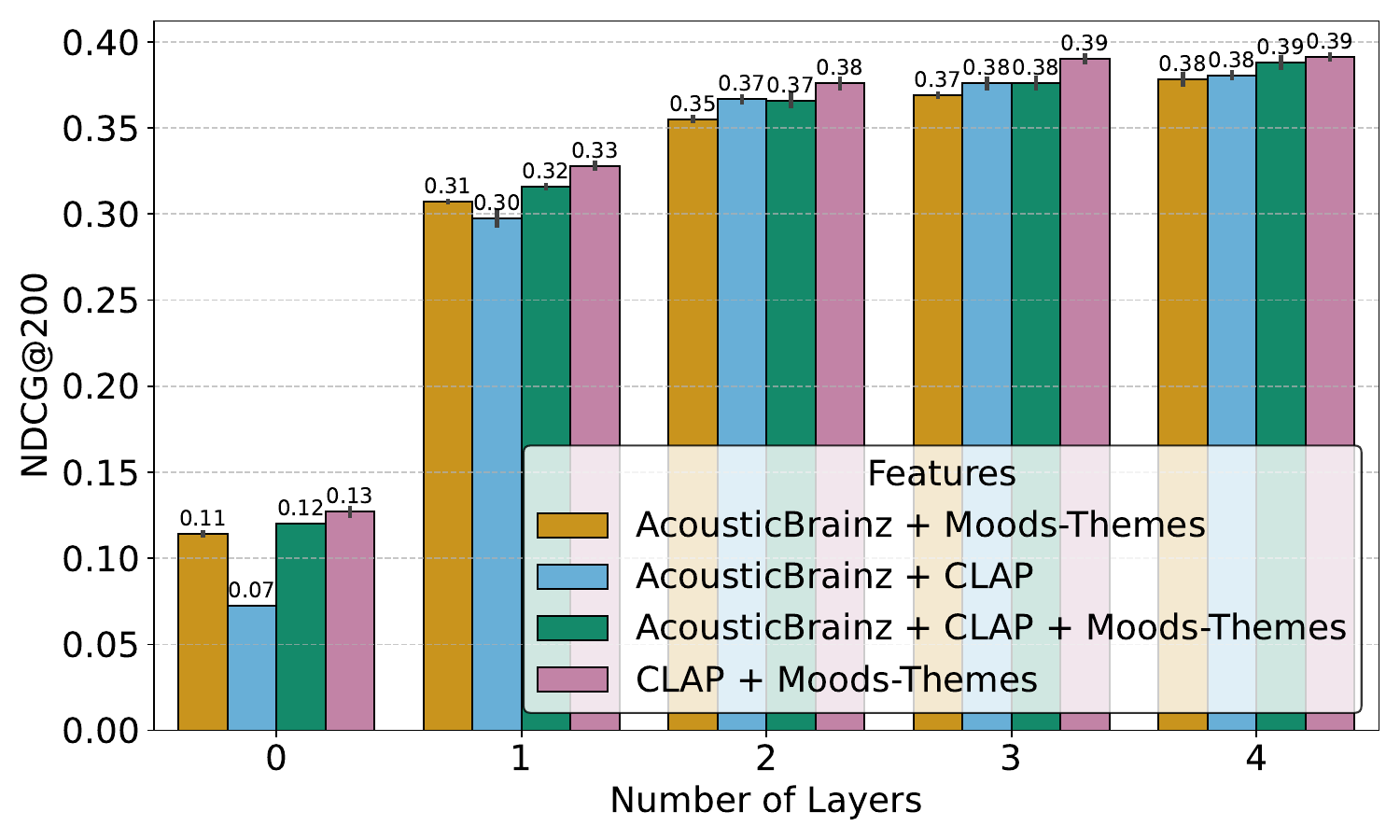}
        \caption{Comparison of various feature combinations. With fewer layers, feature combinations perform better than single features, whereas they perform on par for more layers.}
        \label{fig:results_b}
    \end{subfigure}
    \caption{Comparison of input features used for the artist relationship prediction task. We report the mean performance and indicate the standard deviation over three seeds for each configuration, testing all setups with 0 to 4 GNN layers. The 0-layer configuration serves as the baseline, where no message-passing is performed, and only the input features are used to predict node pairs. }
    \label{fig:results}
\end{figure}

Figure~\ref{fig:results_a} compares the performance of models using random features, AcousticBrainz features, Moods-Themes features, and CLAP features. The baseline model, which does not utilize any graph convolution layers, performs significantly worse than models incorporating graph topology information. Performance generally improves with the addition of more graph layers. Random features consistently underperform, while CLAP features show better results with increased layers in comparison to the others. 
Moods-Themes features perform well without graph layers but only achieve results similar to random features with four layers, indicating that the information they provide can be compensated by knowledge of the neighborhood around an artist. 
Based on these findings, we conclude that CLAP embeddings are effective in enhancing music recommendation tasks and provide information that is missing in other features.

We further compare combinations of CLAP embeddings with other features to assess their effectiveness. Our analysis in Figure~\ref{fig:results_b} reveals that for lower layer numbers, the combination of features can greatly increase performance in comparison to single features (as depicted in Figure~\ref{fig:results_a}). For more layers, the tested feature combinations approach the performance of the model that only uses CLAP features. This could mean that the other features do not provide much additional value for the task or that the information gained from the graph topology is sufficient to compensate for it.  
Overall, feature combinations that include CLAP perform better, while we can see a clear increase of AcousticBrainz + Moods-Themes over the single feature baselines. 

\paragraph{Limitations}
Our experimental evaluation has two main limitations: the potential for model architecture improvements and the limited representation of artists using only one song.

First, regarding model architecture, there is room for enhancement through more advanced techniques, such as distance-weighted sampling, more sophisticated GNN layers, or Graph Transformers. We anticipate these improvements would likely lead to better overall performance. However, our conclusions primarily focus on the relative performance gains of different feature sets. We believe these relative differences would remain consistent even with improved models and training techniques, though absolute performance might increase.

Second, we only use a single song to represent each artist. This approach could introduce variability based on the choice of the song, potentially affecting the performance of the features. A more comprehensive representation involving multiple songs per artist could provide a more robust understanding, but this would require careful consideration of how to aggregate these song embeddings. Additionally, there is potential for exploring different versions of CLAP or other audio embedding models. Nevertheless, the fact that we achieved consistent performance gains even with just one song per artist demonstrates the effectiveness of CLAP embeddings as a viable approach for music recommendation, which was the primary objective of this study.

\section{Conclusion}

In this work, we explored the use of CLAP embeddings as descriptive data for music recommendation systems. Our experiments focused on a graph-based artist-relationship prediction task, comparing the effectiveness of various feature representations, including AcousticBrainz, CLAP, and a combination of both. Our results indicate that models incorporating CLAP embeddings significantly outperform those using traditional features, particularly as the number of graph convolutional layers increases. This highlights the potential of CLAP embeddings to capture rich and relevant information about music, thereby enhancing the performance of music recommendation systems.

%%
%% Define the bibliography file to be used
\bibliography{sample-ceur}

\appendix


% The sources for the ceur-art style are available via
% \begin{itemize}
% \item \href{https://github.com/yamadharma/ceurart}{GitHub},
% % \item \href{https://www.overleaf.com/project/5e76702c4acae70001d3bc87}{Overleaf},
% \item
%   \href{https://www.overleaf.com/latex/templates/template-for-submissions-to-ceur-workshop-proceedings-ceur-ws-dot-org/pkfscdkgkhcq}{Overleaf
%     template}.
% \end{itemize}

\end{document}